\documentclass[amsfonts,amsmath,amssymb,a4paper]{revtex4}

\usepackage{graphicx}
\usepackage{dcolumn}
\usepackage{bm}

\def\cg{\cal G}

\begin{document}

\preprint{Bicocca-FT-02-11}
\title{Recurrence on the average on trees}
\author{Luca Donetti}
\email{Luca.Donetti@mib.infn.it} \affiliation{Dipartimento di Fisica
G. Occhialini, Universit\`a di Milano--Bicocca and INFN, Sezione di
Milano, Piazza delle Scienze 3 - I-20126 Milano, Italy}

\begin{abstract}
  In this paper we show that all infinite trees which have bounded
  coordination and whose surface is negligible with respect to the
  volume in the limit of large distances (so that they can be embedded
  in a finite--dimensional euclidean space) are recurrent on the
  average; this has important consequences about the spontaneous
  symmetry breaking of statistical models defined on such trees.
\end{abstract}

\maketitle

\section{Introduction}
\label{sec:intro}
Random walks, naturally connected to the problem of diffusion on
discrete structures, are also important tools for studying the
properties of inhomogeneous networks. In fact they represent the main
link between the geometrical and the physical properties of graphs: a
walker needs only the information about local connectivity for the
one--step jump probabilities, but for sufficiently long times it
samples the whole underlying structure, so that its long time
statistics reflects the large scale topology which, in turn, is
responsible for many physical properties.

Random walks also represent a connection between mathematics and
physics: these are extensively studied in the mathematical literature
because of their connections with Markov chains, algebraic graph
theory and potential theory (see for example \cite{woess}) while, on
the other side, many physical problems such as the study of
vibrational spectrum \cite{alexorb}, the critical behavior of
statistical models \cite{hhw} or simple models of quantum particles
\cite{tightbind} on discrete networks can be mapped to random walks.
In particular the Type problem, i.e.  whether for a certain graph a
random walker returns to its starting point with probability one ({\em
  recurrent} graph) or has a nonzero chance to escape ({\em transient}
graph), is strictly related to the presence of spontaneous symmetry
breaking of statistical models defined on the same structure
\cite{cassiloc}.


For inhomogeneous structures the local properties can be different
from the average ones, so the Type problem on the average has
been introduced \cite{avgtype}: a graph is said to be {\em recurrent
  on the average} (ROA) or {\em transient on the average} (TOA) on the
basis of the average over all nodes of the returning probability. This
new classification has been shown to be the appropriate one in the
analysis of statistical models: in particular on ROA graphs, classical
$O(n)$ and quantum Heisenberg ferromagnetic spin models cannot show
spontaneous magnetization at any finite temperature \cite{cassiavg};
on the contrary on TOA ones, $O(n)$ models \cite{invmwavg} must have a
non--zero magnetization at finite temperature.

In this paper we show that a whole class of trees, which are defined
{\em bounded} trees in \cite{rndtree}, are recurrent on the average.
This class is composed by all infinite trees that satisfy two
requirements with a natural physical meaning: first the coordination
number of the nodes must be bounded (when the graph is embedded in an
euclidean space, the number of nearest neighbors of a given site has a
geometrical upper bound if the links have bounded length), then the
``surface'' must be negligible with respect to the ``volume'' for
large distances (this is necessary for the existence of the
thermodynamic limit or simply if the graph has to be embedded in a
finite--dimensional Euclidean space). The proof is based on the flow
criterion: this is a powerful tool for the study of the local Type
problem \cite{localflow} and recently it has been extended to the case
on the average \cite{bertacchi}. The result obtained in this paper has
an important consequence that follows from the previously stated
connections between random walk and statistical models defined on
graphs: there cannot be spontaneous symmetry breaking for classical
$O(n)$ and quantum Heisenberg ferromagnetic models on any bounded
tree.

Moreover the recurrence on the average implies that the average
spectral dimension $\bar d_s$ of these trees is always smaller than 2;
this is a first exact result about the average spectral dimension for
bounded trees even if we expect the true upper limit to be $4/3$.
Indeed this has been conjectured (but not proved) in \cite{rndtree2}
and the inequality $\bar d_s \le \frac43$ is satisfied by all known
examples of bounded trees.

Another remark about our result is that there are no corresponding
results concerning the local recurrence: there are indeed examples of
locally transient bounded trees \cite{ntd}.

The outline of the paper is as follows: in next section we recall
the definitions about graph averages, recurrence and transience, and
the flow criterion; then in section~\ref{sec:main} we formally state
and prove our result about the recurrence on the average of bounded
trees.

\section{Definitions}
\label{sec:defin}
Given an infinite graph $\cg$, let us call $G$ the set of vertices
(nodes) and $E(\cg)$ the set of edges (links). For $x \in G$ the
sphere $B(x,r)$ and the spherical shell $S(x,r)$ are defined as:
\begin{gather*} 
  B(x,r)=\{y\in G: d(x,y)\le r \} \\
  S(x,r)=\{y\in G: d(x,y)=r \} 
\end{gather*}
where $d(x,y)$ is the chemical distance on the graph $\cg$.

Now consider a function $f$ defined on $G$; the limit on the average $L_x(f)$
is defined as \cite{bertacchi}:
\begin{equation} \label{eq-limit} 
  L_x(f)= \lim_{r\to \infty} \frac1{|B(x,r)|} \sum_{y\in B(x,r)} f(y)
\end{equation}
If $f=\chi_A$, the characteristic function of a subset $A\subseteq G$,
we write $L_x(A)$ instead of $L_x(\chi_A)$ and call it measure of $A$.
We will consider also the upper and lower limit on the average:
\begin{gather*} 
  \inf L_x(f) = \liminf_{r\to \infty} \frac1{|B(x,r)|} 
  \sum_{y\in B(x,r)} f(y) \\ 
  \sup L_x(f) = \limsup_{r\to \infty} \frac1{|B(x,r)|} 
  \sum_{y\in B(x,r)} f(y)
\end{gather*}
because these always exist while the limit in
equation~(\ref{eq-limit}) may not exist.

\subsection{Transience and recurrence}
Given a graph $\cg$ and a node $x\in G$, consider the probability
$F_x$ for a random walker started from $x$ to ever return to the
starting vertex: $\cg$ is said to be {\em (locally) transient} if $F_x < 1$
and {\em (locally) recurrent} if $F_x = 1$.

This classification is a property of the graph and does not depend on
the choice of starting node $x$ because it is easy to prove, using
standard Markov chain properties, that if $F_x=1 $ ($F_x<1 $) for any
node $x \in G$ then $F_y = 1$ ($F_y<1 $) for every other node $y \in
G$.

\subsection{Transience and recurrence on the average}
A graph is said to be {\em transient on the average} if, for $x\in G$
\begin{equation*}
 \inf L_x(F) < 1 
\end{equation*}
and {\em recurrent on the average} (ROA) if, on the contrary,
\begin{equation*}
 \inf L_x(F) = 1 
\end{equation*}

Important properties of the classification on the average are
\cite{bertacchi}:
\begin{itemize}
\item if there exists $y \in G$ satisfying
\begin{equation*}
  \sup_r \frac{|S(y,r+1)|}{|B(y,r)|} < +\infty
\end{equation*}
then $\inf L_x(F)=1$ for all $x \in G$ or $\inf L_x(F)<1$ for all $x
\in G$ (the classification does not depend on $x$)
\item local recurrence implies recurrence on the average, while this is
  not true for transience.
\end{itemize}

\subsection{Flow criterion}
If we give an orientation to the links of a graph, i.e. for every 
$e\in E(\cg)$ we write $e = (e^-,e^+)$
with $e^-,e^+ \in G$, a {\em flow from $x_0$ to infinity with input
  $i_0$} is a function $u$ defined on $E(\cg)$ such that the
``current'' is conserved at every vertex $x$:
\begin{equation*} 
  \sum_{e:e^-=x}u(e) - \sum_{e:e^+=x}u(e) = \delta_{x,x_0}i_0
\end{equation*}
and its {\em energy} is
\begin{equation} \label{eq:energy} 
  \langle u, u\rangle = \sum_{e \in E(\cg)} u(e)^2
\end{equation}

The existence of finite energy flow is strictly related to transience, as
expressed by the following:
\begin{itemize}
\item A graph $\cg$ is (locally) transient if and only if there exists
  $x \in G$ such that it is possible to find a finite energy flow
  $u^x$ with non-zero input from $x$ to infinity \cite{localflow};
\item A graph $\cg$ with a bounded coordination is TOA if and only if
  there exists $A \subseteq G$ such that $\sup L_o(A) > 0$ and for
  every $x \in A$ there is a finite energy flow $u^x$ from $x$ to
  $\infty$ with non-zero input and $\sup_{x \in A} \langle u^x,
  u^x\rangle < +\infty$ (theorem 3.10 in \cite{bertacchi}).
\end{itemize}

\section{Main result}
\label{sec:main}

We prove that: \\
\noindent
{\em  If $\cg$ is an infinite tree with bounded coordination and
  there exists $x_0 \in G$ such that
\begin{equation*}
  \lim_{r \to \infty} \frac{|S(x_0,r)|}{|B(x_0,r)|} = 0 
\end{equation*}
then $\cg$ is ROA.}

The outline of the proof is the following: first we define the
``backbone'' of $\cg$ and the set $V$ of the backbone ``branching''
nodes and we show that it has measure zero; then we prove that for every
set $A \subseteq G$ such that $\sup L_{x_0}(A) > 0$ and for every $n
\in \mathbb{N}$ there exists $x \in A$ such that every flow $u^x$ from
$x$ to infinity with input $i_x$ has energy $\langle u^x, u^x\rangle >
\frac{i_x^2}{2}n$, so that $\sup_{x \in A} \langle u^x, u^x\rangle$ is
infinite and the graph cannot be TOA.
 
\subsection{Proof}
\begin{figure}[t]
  \includegraphics[width=0.4\textwidth]{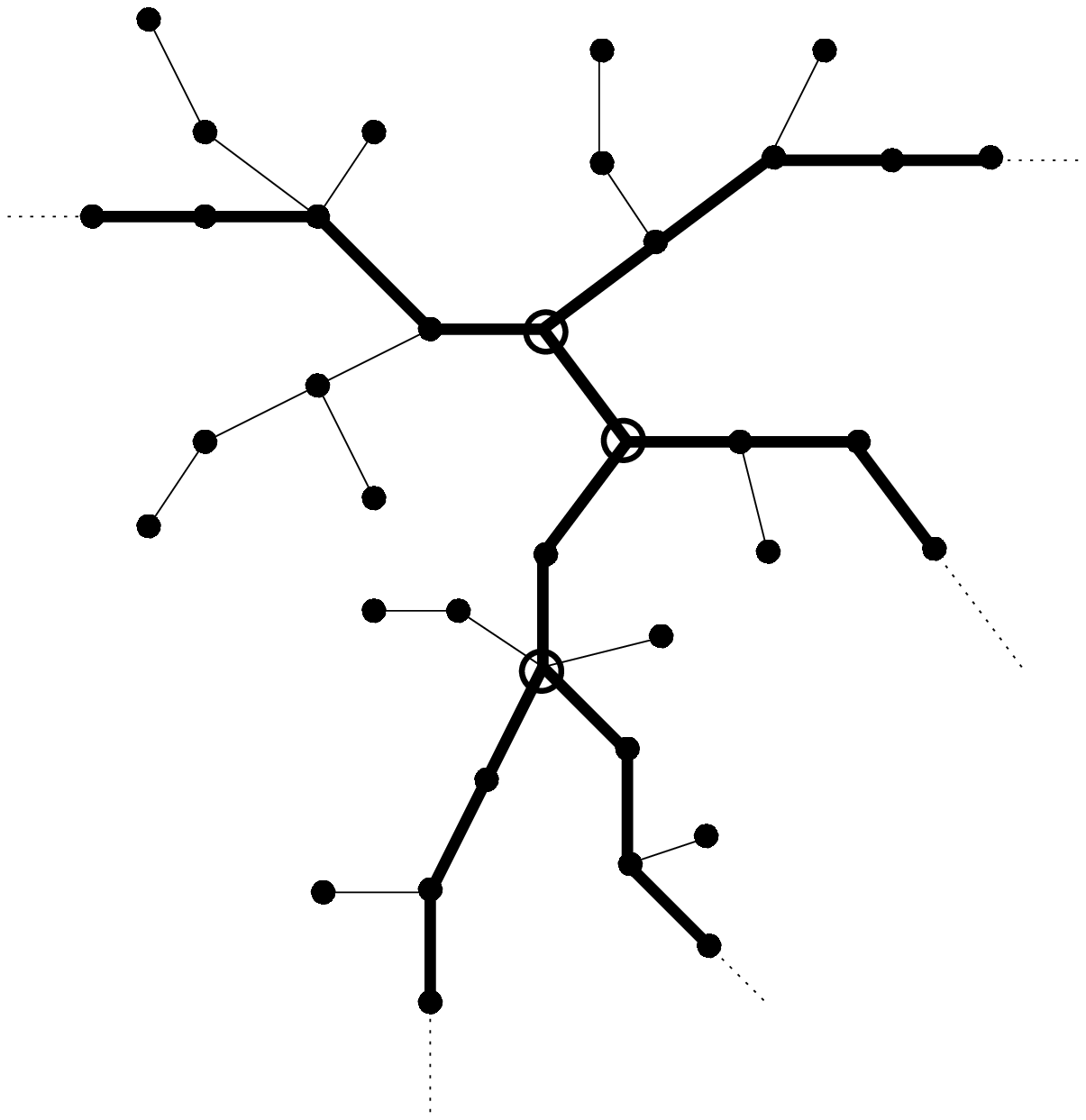}
\caption{The tick line shows the backbone of the tree, the circle 
  its branching.}
\label{backb}
\end{figure}

Consider the subgraph $\cal R$ (with vertex set $R$) obtained from
$\cg$ by deleting all vertices and links belonging only to finite
branches and call it the backbone of $\cg$ (fig~\ref{backb}): it can
be obtained by recursively deleting all nodes with coordination one
together with the links originating from them.
  
If $R$ is empty (for example if $\cg$ is a half infinite
one--dimensional chain, possibly with finite trees attached to some or
all nodes) then $\cg$ is recurrent (and therefore ROA). In this case,
in fact, for every node $x\in G$ there is only one path that leads to
infinity: because of current conservation $|u^x(e)|$ must be finite
and constant on the edges $e$ belonging to this path and zero on every
other edge and this implies that $\langle u^x, u^x\rangle = +\infty$.

If, on the other hand, $R$ is non--empty let us call ``branching'' the
nodes $x\in R$ whose coordination in $\cal R$ is greater than 2; these
vertices are important because they are the only ones in which a
current can split so that the flow energy can be reduced. Now define
$V$ the set of branching nodes, $B_V(x_0,r)$ the set of branching
nodes inside a ball of radius $r$ and center $x_0$ and $S_R(x_0,r)$
the vertices of $S(x_0,r)$ belonging to $R$:
\begin{gather*} 
  V = \{x:x\in R, z^{\cal R}_x > 2\} \\
  B_V(x_0,r) = B(x_0,r) \cap V \\
  S_R(x_0,r) = S(x_0,r) \cap R
\end{gather*}

If there are no branching vertices ($V$ is empty) $\cal R$ is a linear
chain (every node has coordination two) and the flow from any vertex
in $G$ has infinite energy so that $\cg$ is recursive (and therefore
ROA).  On the other hand even if $V$ is non--empty, it is a zero measure
set in $G$ because
\begin{eqnarray*}
|S(x_0,r)| & \ge & |S_R(x_0,r)| = 2 + \sum_{x\in B_V(x_0,r)}(z^{\cal R}_x-2) \\
         & \ge & 2 + |B_V(x_0,r)| \\
         &  >  & |B_V(x_0,r)|
\end{eqnarray*}
so that
\begin{equation*} 
 L_{x_0}(V) =  \lim_{r\to \infty}  \frac1{|B(x_0,r)|}  
 \sum_{x\in B(x_0,r)} \chi_V(x) =
 \lim_{r\to \infty}\frac{|B_V(x_0,r)|}{|B(x_0,r)|} \le 
 \lim_{r\to \infty}\frac{|S(x_0,r)|}{|B(x_0,r)|} = 0 
\end{equation*}

Now consider a subset $A \subseteq G$ such that $\sup L_{x_0}(A) > 0$, it
is easy to see that for every $n$ there exists $x \in A$ such that
$d(x,V) > n$. In fact, suppose that there exists $\bar n$
such that for every $x\in A$, $d(x,V) \le \bar n$, then
\begin{equation*}
  A \subseteq M_{\bar n} = \{ x\in G: d(x,V) \le \bar n\} 
\end{equation*}
but since $V$ has zero measure and $\cg$ has bounded coordination,
$M_{\bar n}$ has zero measure for every $\bar n \in \mathbb N$ (lemma
4.8 in \cite{bertacchi}) and so $L_o(A) = 0$ contradicting the
hypothesis.

Then choose a distance $n \in \mathbb N$ and take $x \in G$ such that
$d(x,V) > n$; there are two distinct cases: $x \in R$ or $x \notin R$.
If $x \in R$, it belongs to finite linear chain in $\cal R$ between
two nodes $v_1,v_2 \in V$ or to a semi--infinite chain starting from
$v_1\in V$.  In the first case let $n_1 = d(x,v_1)$ and $n_2=d(x,v_2)$
with, by hypothesis, $n_1>n$ and $n_2>n$.  Clearly the energy of a
flow $u_x$ from $x$ decreases if we restrict the sum in equation
(\ref{eq:energy}) to the linear chain between $v_1$ and $v_2$ and all
the ``restricted''energy of the flows can be parametrized by the
fractions $t$ and $1-t$ of current directed toward the nodes $v_1$ and
$v_2$, respectively:
\begin{eqnarray*}
  \frac{\langle u^x, u^x\rangle}{i_x^2} & > & 
    \sup_{0\le t \le 1} n_1 t^2 + n_2 (1-t)^2 =
    \frac{n_1 n_2}{n_1+n_2} \\ 
  & \ge & \frac{n^2}{2 n} = \frac{n}{2}
\end{eqnarray*}
On the other hand if $x$ belongs to a semi--infinite chain, in order
to have a finite energy flow the $|u^x(e)|$ must be constant in the
path between $x$ and $v_1$ and zero in the other (infinite) part of
the semi--infinite chain; so we have
\begin{equation}
 \frac{\langle u^x, u^x\rangle}{i_x^2} > n_1 > \frac{n_1}2 > \frac{n}2
\end{equation}
If $x\notin R$ let $y$ be the vertex in $R$ with minimum distance from
$x$ and $n_1=d(x,y)$. In this case $|u^x(e)|$ must be constant for
every edge $e$ belonging to the path connecting $x$ to $y$ and then
the current can split as in the previous case.
\begin{eqnarray}
 \frac{\langle u^x, u^x\rangle}{i_x^2} & = &
        n_1 + \frac{\langle u^y, u^y\rangle}{i_x^2} \nonumber\\
 & > & n_1 + \frac{n-n_1}{2}  \nonumber\\
 & > &  \frac{n}{2} 
\end{eqnarray}
Collecting all these results, we can say that for every $x\in G$ such
that $d(x,V) > n$
\begin{equation*}
  \frac{\langle u^x, u^x\rangle}{i_x^2} > \frac{n}{2}  
\end{equation*}
This implies that for every subset $A \subseteq G$ such that $\sup
L_{x_0}(A) > 0$, it must be $sup_{x \in A} \langle u^x, u^x\rangle =
+\infty$, so the graph $\cg$ is not TOA and must be ROA.

\begin{acknowledgments}
  I would like to thank C. Destri, D. Cassi and R. Burioni for
  useful discussions and suggestions.
\end{acknowledgments}

\end{document}